\documentclass[prb,preprint,showpacs,preprintnumbers,amsmath,amssymb]{revtex4}
\usepackage{graphics}
\begin{document}

\title{Inducement and suppression of Coulomb effects in elastic
2D-2D electron tunnelling in a quantizing magnetic field}

\author{{\em V.~G.~Popov}$^1$, \fbox{Yu.~V.~Dubrovskii}$^1$,
J.-C.~Portal$^2$}

\affiliation{$^{1}$~Institute of Microelectronics Technology of RAS, Chernogolovka 142432, Russia.\\
$^{2}$~INSA, F31077 Toulouse Cedex 4, France.\\ $^3$ Grenoble High
Magnetic Field Laboratory, MPI-CNRS, BP166 38042 Grenoble Cedex 9,
France.}

\date{\today}

\begin{abstract}
{Tunnelling between two-dimensional electron systems has been
studied in the magnetic field perpendicular to the systems planes.
The satellite conductance peaks of the main resonance have been
observed due to the electron tunnelling assisted by the elastic
scattering on impurities in the barrier layer. These peaks are
shown to shift to the higher voltage due to the Coulomb pseudogap
in the intermediate fields. In the high magnetic fields the
pseudogap shift is disappeared.}
\end{abstract}

\maketitle

\section*{Introduction}

Coherent 2D-2D tunnelling is well known to have a resonance when
subband energies $E_{01}$ and $E_{02}$ coincide in both
two-dimensional electron systems (2DESs). Quantizing a lateral
motion of electrons a normal magnetic field sharpens the coherent
resonance and produces a series of satellite resonances originated
from tunnelling between Landau levels (LL) with different numbers
assisted the elastic scattering on impurities~[1]. The condition
of the elastic 2D-2D tunnelling is the following:
$E_{01}-E_{02}=k\hbar \omega_{c}$. Here $k$ is an integer number,
$\omega_{c}$ is the cyclotron frequency. In other words in tunnel
spectra the nearest satellites are separated by a voltage interval
$V_{c}=\hbar \omega_{c}/e$. This single particle picture ignores
many-body effect such as the Coulomb pseudogap (CP) that provides
an additional voltage shift of the coherent resonance~[2]. The
pseudogap is originated from relaxation processes induced by the
tunnelling electrons. In other words the tunnelling electron
should have an additional energy to spend it on the relaxation
therefore tunnelling is suppressed at low energies. This scenario
is valid while the tunnelling time is very short in compare with
the relaxation times. The interesting question is what happens if
the relaxation will take part in tunnelling of electrons. To
clarify the question it is suitable to investigate processes
providing peak features in the tunnel spectra. The features of the
elastic 2D-2D tunnelling are such kind of features. Here the
Coulomb effects is reported to be revealed and studied in elastic
2D-2D electron tunnelling. In particular an inducement and a
suppression of the Coulomb effects have been observed in a
quantizing magnetic field.

\begin{figure}
\includegraphics{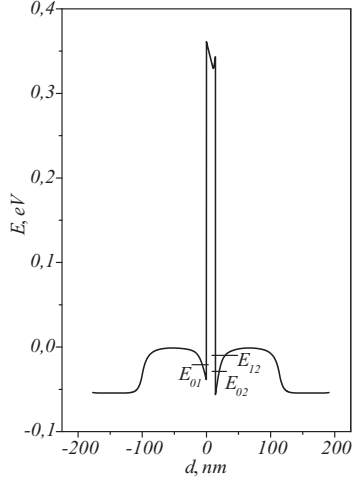}\caption{The potential
profile of the conductance-band bottom of the heterostructure
under investigation with the quantum levels calculated
self-consistently.}\label{profile}
\end{figure}

\section*{Experiment}
\begin{figure}[t]
\includegraphics{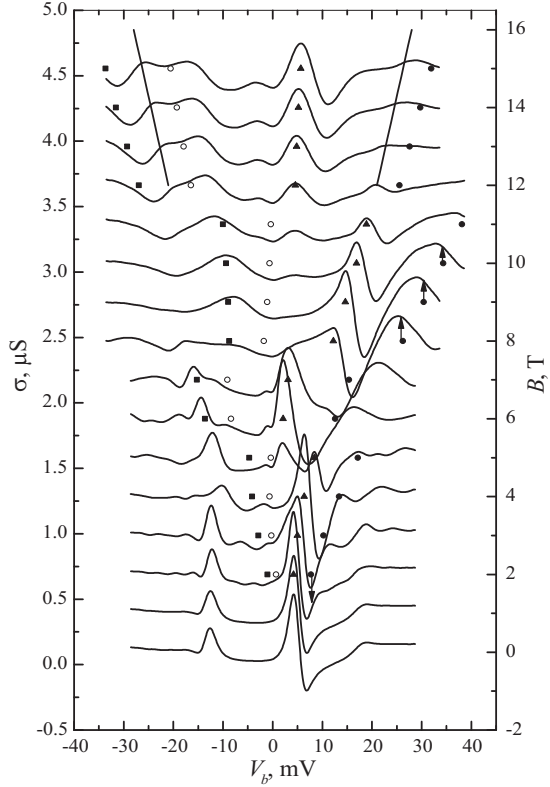}\caption{Tunnel
conductance-voltage dependencies at different magnetic fields in
combination with the peaks positions magnetic dependencies. The
curves are shifted to correspond magnetic field scale B, i.e. with
step being proportional to the field that.}\label{fig}
\end{figure}

The investigated tunnel diodes represented columns wet-etched in a
single barrier heterostructure of GaAs/ \\
Al$_{0.3}$Ga$_{0.7}$As/GaAs type. The barrier layer was of 20\,nm
thickness and doped with Si at the middle. Due to the donor
ionization the 2DESs accumulate on both side of the tunnel barrier
in the undoped spacer layers of 70\,nm thickness. The resistance
of the spacers was quite low in compare with the barrier one. This
provided common Fermi levels in 2DESs and adjacent contact
n$^+$-GaAs regions. The potential profile of the heterostructure
is shown in Figure~\ref{profile}. The symmetry of the structure is
supposed to be a result of the dopants diffusion during the
crystal growth. The conductance-voltage dependencies are shown in
Figure~\ref{fig} at different magnetic fields. At zero field the
conductance peaks are the coherent resonant those. In particular
the peak at $V_b = 3$\,mV corresponds to the 0-0 resonance, i.e.
$E_{01} = E_{02}$ and peak at $V_b = -12$\,mV takes place when
$E_{01} = E_{12}$. The electron concentrations were determined
from magneto-oscillations of the tunnel current at low bias
voltages and from the voltage position of the current peak at zero
magnetic field. They are $n_1 = 3 \times 10^{11}$\,cm$^{-2}$ and
$n_2 = 5 \times 10^{11}$\,cm$^{-2}$ the mobilities can be also
estimated as $\mu_{1,2} \approx 50000$\,cm$^2/$Vs~[3]. The I-V
curves of the tunnel diodes demonstrate a nonmonotonous magnetic
dependence of the coherent-resonance position. The voltage
dependencies of the tunnel conductance have pronounced satellite
resonances or the conductance peaks shown in Figure~\ref{fig}. The
satellite peaks also have nonmonotonous magnetic dependencies
resembled the coherent resonance one (see Fig.~\ref{fig}).

\section*{Discussion}
The triangles in Figure~\ref{fig} follow the coherent peak
position. The circles show the expected values for the first
satellite peaks separated from coherent one at $V_{\pm s}= \pm
\hbar \omega_c / e$. One can see that the expected positions are
in quite good agreement with experimental ones at the positive
voltage polarity (filled circles in Fig.~\ref{fig}) and in
disagreement with those at negative polarity (empty circles in
Fig.~\ref{fig}) at least for the magnetic field range $B\in$(6\,T;
11\,T). It is interesting to note that the deviation can be
cancelled with an additional voltage shift depended upon magnetic
field. So if one supposes $V_{-s}\approx -1.5 \hbar \omega_c / e$
the data coincidence will be better (see. squares at the negative
voltage in Fig.~\ref{fig}). Such shift can be explained by the CP.
In this case the positive satellite shifts together with the
coherent peak and the energetic interval between them remains the
same $eV_{+s} = \hbar \omega_c$. As for the negative satellite it
is experienced the reverse shift and the energetic distance is
increased on the double value of the CP $eV_{-s}=-\hbar \omega_c -
2\Delta_C$. The value of the CP can be estimated from the splitted
coherent peak at very high magnetic fields . At these fields both
2DESs have the only one populated LL. Under this ultraquantum
limit the resonant voltage should be zero~[3]. In this case
deviation originates only from the CP. Hence one can easy estimate
the CP as a value of the deviation, i.e. $\Delta_C \approx 0.3
\hbar \omega_c$. Thus one can justify the large value of $V_{-s}$.
The next interesting feature appears at high magnetic fields $B >
11$\,T where both the empty circles and the squares have lack to
describe the satellite peaks positions. Moreover each peak has
splited on a strongly and a weakly field depended ones. The
strongly field-depended peaks follow to the cyclotron lines, i.e.
$eV_{\pm p} = \pm \hbar \omega_c $ (see solid lines in
Fig.~\ref{fig}). They can be considered as elastic satellites
without the CP shift. The weakly depended peaks can be assigned to
resonant tunnelling between the ground and the first excited
subbands. Thus we can interpret the experimental data as a
suppression of the CP effect on elastic peaks. Such suppression
can be expected because, when the LL energy exceeds the first
excited subband one, the intensive inter-subband scattering can
decrease an electron life-time on the LL, i.e. the relaxations
becomes faster, and thus decrease the role of the Coulomb effects.

\section*{Conclusions}

The Coulomb pseudogap has been found  to cause an additional
voltage shift of the conductance peaks originated from 2D-2D
tunnelling assisted by the elastic scattering. The disappearance
of the shift has been also observed in the strong field when
$\hbar \omega_c > E_{12}-E_{02}$ (see Fig.~\ref{profile}). The
appearance of the CP effects means that elastic tunnelling is
still quite fast in compare with lateral relaxation. The effect
disappearance takes place probably due to the additional
relaxation process, i.e. inter-subband scattering, becomes
significant.

\begin{acknowledgments}
This work has been supported in part by the RFBR (grant
07-02-00487), INTAS (grant YSF 05-109-4786), "Russian Science
Support Foundation", and RAS program "Quantum nanostructures".
\end{acknowledgments}

\end{document}